\documentclass[aps,prl,twocolumn,groupedaddress]{revtex4-2}

\usepackage{amssymb}
\usepackage{amsmath}
\usepackage{graphicx}
\usepackage{enumerate}
\usepackage{mathtools}
\usepackage{color}
\usepackage{bbold}
\usepackage{subfigure}
\usepackage{slashed}
\usepackage{xcolor}
\usepackage{bm}

\begin{document}


\title{Collisional flavor instabilities of supernova neutrinos}


\author{Lucas Johns}
\email[]{NASA Einstein Fellow (ljohns@berkeley.edu)}
\affiliation{Departments of Astronomy and Physics, University of California, Berkeley, CA 94720, USA}

\begin{abstract}
A lingering mystery in core-collapse supernova theory is how collective neutrino oscillations affect the dynamics. All previously identified flavor instabilities, some of which might make the effects considerable, are essentially collisionless phenomena. Here it is shown that collisional instabilities exist as well. They are associated with asymmetries between the neutrino and antineutrino interaction rates, are possibly prevalent deep inside supernovae, and pose an unusual instance of decoherent interactions with a thermal environment causing the sustained growth of quantum coherence. 
\end{abstract}

\maketitle

\textit{Introduction.}---Experiments and observations, in the real world and on computers, have confirmed time and again that neutrinos oscillate and are critical to core-collapse supernova explosions. The jury is still out, however, on how neutrinos oscillate \textit{in} supernovae \cite{mirizzi2016}. 

One way to approach the problem is to take the output of hydrodynamic simulations as a starting point. State-of-the-art simulations evolve neutrinos classically in the sense that the particles are non-oscillating and therefore never develop quantum flavor coherence. To figure out whether oscillations engender large effects on the supernova dynamics, the first question to ask is whether such classical flavor-field solutions support collective instabilities \cite{banerjee2011,izaguirre2017,capozzi2017}. If they do, then supernova neutrino transport may not be meeting the standards it aspires to after all.

The current body of evidence points to \textit{fast instabilities} \cite{sawyer2016, chakraborty2016, tamborra2020} as a likely culprit in steering post-shock flavor fields off their classical courses. The name refers to their growth rates being proportional to $G_F n_\nu$, which translates to a timescale on the order of nanoseconds at distances a few tens of kilometers from the supernova center. They are to be distinguished from \textit{slow instabilities} \cite{duan2010}, which grow at $\sqrt{G_F n_\nu \omega}$ rates ($\omega$ is the vacuum oscillation frequency) and appear to pose less of a threat interior to the stalled shock than fast instabilities do. The fact that $G_F n_\nu$ dictates the scales of both sets of phenomena reveals that they really fall within a single class, both being driven by nonlinear refraction due to neutrino--neutrino forward scattering. One could fairly say that this class has defined the subject matter of the field of collective neutrino oscillations.

In this paper the existence of another class of collective phenomena is established. \textit{Collisional instabilities}, while enabled by nonlinear refraction, are a $G_F^2$ effect, operating on scales set by the neutrino collision rate. Like fast instabilities, they may jeopardize the validity of classical flavor fields behind the supernova shock.

The idea that collisions are responsible for instability is seemingly paradoxical. Neutrino transport is described by the equation of motion
\begin{equation}
i \left( \partial_t + \mathbf{v} \cdot \partial_{\mathbf{x}} + \dot{\mathbf{p}} \cdot \partial_{\mathbf{p}} \right) \rho = \left[ H, \rho \right] + i C, \label{qke}
\end{equation}
where $\rho$ is the flavor density matrix, $H$ is the Hamiltonian, and $iC$ is the collision term. Suppose that a classical flavor field---the output of a radiation-hydrodynamics simulation---is found to be stable to collisionless instabilities but unstable to collisional ones. The implication is that this flavor field solves Eq.~\eqref{qke} with $H = 0$ and $C \neq 0$, approximately solves the equation with $H \neq 0$ and $C = 0$ (and appropriate boundary conditions), but \textit{fails} to be a reliable solution when the terms are nonzero simultaneously. This circumstance contravenes the intuition, built up over decades, that the effect of collisional decoherence is always to damp or suppress oscillations.

In the following, collisional instability is exhibited by directly manipulating the nonlinear equations of motion, by numerically evolving them using parameters representative of a real supernova, and by analytically solving the dispersion relation of the linearized system. The likely prevalence and consequences are discussed, as are possible extensions of the analysis. It is shown, in this last connection, that fast flavor conversion can incite the growth of collisional instabilities.

\textit{Collisional instability.}---To demonstrate the phenomenon, we first consider an isotropic and homogeneous neutrino system, representing a small region of the flavor field in a core-collapse supernova. For collisions we adopt the relaxation-time approximation \cite{mckellar1994, dolgov2001, hannestad2015, johns2019b}, applying it separately to absorption and emission (\textit{AE}), charged-current scattering (\textit{CC}), and neutral-current scattering (\textit{NC}):
\begin{align}
i C = \left\lbrace i \Gamma_{AE}, \rho_{AE} - \rho \right\rbrace &+ \left\lbrace i \Gamma_{CC}, \rho_{CC} - \rho \right\rbrace \notag \\
& + \left\lbrace i \Gamma_{NC}, \rho_{NC} - \rho \right\rbrace,
\end{align}
where $\Gamma_P = \textrm{diag} \left( \Gamma_e^P / 2, \Gamma_x^P / 2 \right)$ for each process type $P$, with $\Gamma_\alpha^P$ being the rate for flavor $\alpha = e, x$. (See Refs.~\cite{richers2019,capozzi2019,shalgar2021,martin2021} for other recent studies of collisions, none of which concern collisional instabilities.) Note that subscript $P$ is used for the matrices. After the next paragraph, only the rates themselves will appear, with $P$ written as a superscript. Note also that \textit{CC} (\textit{NC}) refers, more precisely, to flavor-resolving (flavor-blind) interactions that preserve the number of neutrinos. A single process, like electron scattering, can contribute to both. Any process that changes the number of neutrinos, regardless of whether it goes through a charged or neutral current, contributes to \textit{AE}.

The terms are separated out in this way because they relax the system differently. Ignoring feedback from flavor conversion, $\Gamma_{AE}$ returns the system to the classical equilibrium set by the composition of the environment; $\Gamma_{CC}$ pushes the system toward kinetic equilibrium; and $\Gamma_{NC}$, because it leaves flavor coherence intact during interactions, redistributes flavor states over momentum. In an isotropic and monochromatic setting, $\Gamma_{NC}$ has no effect.

\begin{figure}
\centering
\includegraphics[width=.43\textwidth]{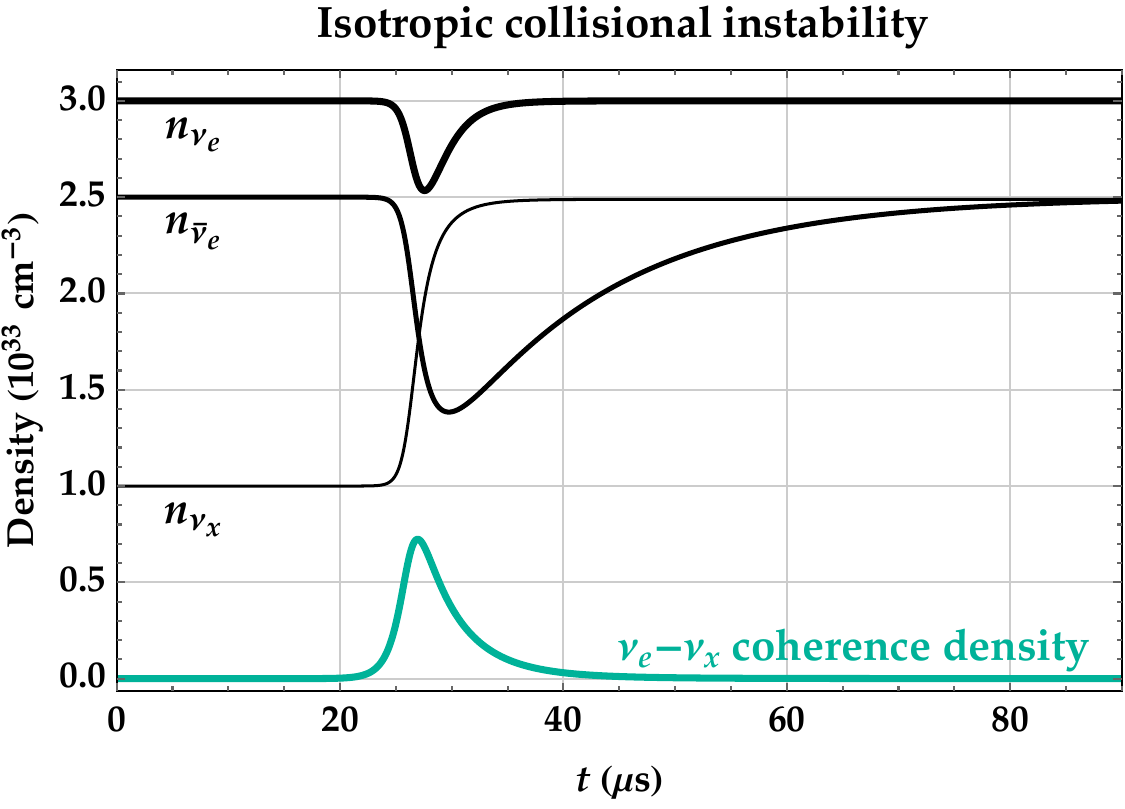}
\caption{Collisionally unstable evolution in an isotropic calculation: $n_{\nu_e}$ (thick black curve), $n_{\bar{\nu}_e}$ (medium), $n_{\nu_x} = n_{\bar{\nu}_x}$ (thin), and neutrino coherence density $| \mathbf{P}_T | / 2$ (teal). For comparison, virtually no flavor conversion or coherence development takes place under collisionless conditions or when $\Gamma$ and $\bar{\Gamma}$ are artificially equated. In the case shown, \textit{de}coherent interactions drive the \textit{growth} of flavor coherence.}
\label{iso}
\end{figure}

Decomposing the density matrix using $\rho = \left( P_0 + \mathbf{P} \cdot \boldsymbol{\sigma} \right) / 2$, with neutrino-number scalar $P_0$ and polarization vector $\mathbf{P}$, the equations of motion become
\begin{align}
&\dot{\mathbf{P}} = \omega \mathbf{B} \times \mathbf{P} + \mu \left( \mathbf{P} - \mathbf{\bar{P}} \right) \times \mathbf{P} - \Gamma_+^{CC}\mathbf{P}_T \notag \\
&\hspace{.25 in} + \Gamma_+^{AE} \left( \mathbf{P}^{AE} - \mathbf{P} \right) + \Gamma_-^{AE} \left( P_0^{AE} - P_0 \right) \mathbf{z} \notag \\
&\dot{P}_0 = \Gamma_+^{AE} \left( P_0^{AE} - P_0 \right) + \Gamma_-^{AE} \left( P_z^{AE} - P_z \right), \label{polveceoms}
\end{align}
where $\Gamma_\pm^{P} = \left( \Gamma_e^{P} \pm \Gamma_x^{P} \right) / 2$. The antineutrino equation is obtained by sending the vacuum Hamiltonian vector $\omega \mathbf{B} \rightarrow - \omega \mathbf{B}$ and putting bars over all rates and vectors except those in the factor $\mathbf{P} - \mathbf{\bar{P}}$. The matter potential $\lambda = \sqrt{2} G_F n_e$ has been dropped because it does not affect stability \cite{hannestad2006}. With the chosen convention, $\mu = \sqrt{2} G_F$.

Although absorption and emission rates are flavor-dependent, the emergence of instabilities is more transparent (and not fundamentally changed) if we take $\Gamma_e^{AE} = \Gamma_x^{AE}$. The total number densities of neutrinos and antineutrinos are then conserved. Letting $\Gamma^P = \Gamma_+^P$ and switching to the sum and difference vectors $\mathbf{S} = \mathbf{P} + \mathbf{\bar{P}}$ and $\mathbf{D} = \mathbf{P} - \mathbf{\bar{P}}$, we have
\begin{widetext}
\begin{gather}
\dot{\mathbf{S}} = \omega \mathbf{B} \times \mathbf{D}  + \mu \mathbf{D} \times \mathbf{S} + \frac{\Gamma^{AE} + \bar{\Gamma}^{AE}}{2} \left( \mathbf{S}^{AE} - \mathbf{S} \right) + \frac{\Gamma^{AE} - \bar{\Gamma}^{AE}}{2} \left( \mathbf{D}^{AE} - \mathbf{D} \right) - \frac{\Gamma^{CC} + \bar{\Gamma}^{CC}}{2} \mathbf{S}_T - \frac{\Gamma^{CC} - \bar{\Gamma}^{CC}}{2} \mathbf{D}_T \notag \\
\dot{\mathbf{D}} = \omega \mathbf{B} \times \mathbf{S} + \frac{\Gamma^{AE} - \bar{\Gamma}^{AE}}{2} \left( \mathbf{S}^{AE} - \mathbf{S} \right) + \frac{\Gamma^{AE} + \bar{\Gamma}^{AE}}{2} \left( \mathbf{D}^{AE} - \mathbf{D} \right) - \frac{\Gamma^{CC} - \bar{\Gamma}^{CC}}{2} \mathbf{S}_T - \frac{\Gamma^{CC} + \bar{\Gamma}^{CC}}{2} \mathbf{D}_T. \label{sdeoms}
\end{gather}
\end{widetext}
Subscript $T$ indicates that only the part of the vector transverse to the flavor axis is being considered. At this point it begins to become clear how collisions might do more than simply decohere flavor states: the $\Gamma - \bar{\Gamma}$ terms couple $\mathbf{S}$ and $\mathbf{D}$ to one another.

From here on we let $\Gamma = \Gamma^{AE} + \Gamma^{CC}$. Now suppose that $\mu \gg \omega, \Gamma, \bar{\Gamma}$. For a system that does not support the bipolar instability, the salient effect of the oscillation terms is to cause $\mathbf{S}$ and $\mathbf{D}$ to undergo synchronized motion around $\mathbf{B}$ \cite{hannestad2006, johns2018}. In a dense matter background, the vectors remain close to the flavor axis and, to a first approximation, the oscillation terms can simply be dropped. Assuming that $| (\Gamma - \bar{\Gamma}) \mathbf{S} | \gtrsim |(\Gamma + \bar{\Gamma}) \mathbf{D} |$, we obtain
\begin{equation}
\ddot{\mathbf{S}}_T + \frac{\Gamma + \bar{\Gamma}}{2} \dot{\mathbf{S}}_T - \left( \frac{\Gamma - \bar{\Gamma}}{2} \right)^2 \mathbf{S}_T \cong 0. \label{steqtn}
\end{equation}
Solutions are exponential, and $\Gamma \neq \bar{\Gamma}$ is required for one of them to be growing.

An instability criterion follows from the assumption that $| (\Gamma - \bar{\Gamma}) \mathbf{S} | \gtrsim |(\Gamma + \bar{\Gamma}) \mathbf{D} |$. In a situation with $\Gamma > \bar{\Gamma}$ and a number-density hierarchy $n_{\nu_e} > n_{\bar{\nu}_e} > n_{\nu_x} \approx n_{\bar{\nu}_x}$, instability is predicted for
\begin{equation}
\mathcal{R} \equiv \frac{n_{\bar{\nu}_e} - n_{\bar{\nu}_x}}{n_{\nu_e} - n_{\nu_x}} \gtrsim \frac{\bar{\Gamma}}{\Gamma} \equiv \mathcal{R}_\textrm{crit}. \label{criterion}
\end{equation}
Numerical tests support the accuracy of this criterion. As $\mathcal{R}$ decreases toward $\mathcal{R}_\textrm{crit}$, the time elapsed before instability sets in grows longer. Below $\mathcal{R}_\textrm{crit}$ the instability has apparently vanished.

The system admits of another collisional instability. Synchronization of $\mathbf{S}$ and $\mathbf{D}$ implies that
\begin{equation}
\dot{\mathbf{D}}_T \cong \left( \pm \frac{\Gamma - \bar{\Gamma}}{2} \frac{ \left| \mathbf{S} \right|}{\left| \mathbf{D} \right|} - \frac{\Gamma + \bar{\Gamma}}{2} \right) \mathbf{D}_T, \label{dteqtn}
\end{equation}
with the upper (lower) sign corresponding to $\mathcal{R} > 1$ ($\mathcal{R} < 1$). If $\mathcal{R} < 1$ and $\bar{\Gamma} > \Gamma$, or if $\mathcal{R} > 1$ and $\bar{\Gamma} < \Gamma$, then an exponentially growing solution is possible, but it is less likely to be of relevance to supernovae.

\textit{An illustrative calculation.}---Figure~\ref{iso} presents the numerical solution of Eqs.~\eqref{polveceoms} using parameters motivated by realistic conditions inside a core-collapse supernova. To be definite, the chosen values emulate those found at, say, a post-bounce time of $\sim$ 200~ms and a radius of $\sim$ 40~km: namely, a density of $10^{12}$~g/cm$^3$, a temperature $T = 7$~MeV, an electron chemical potential $\mu_e = 20$~MeV, and neutrino number densities $n_{\nu_e} = 3 \times 10^{33}$~cm$^{-3}$, $n_{\bar{\nu}_e} = 2.5 \times 10^{33}$~cm$^{-3}$, and $n_{\nu_x} = n_{\bar{\nu}_x} = 1 \times 10^{33}$~cm$^{-3}$. Since the calculation is monochromatic, an energy $E_\nu = 20$~MeV is used for all (anti)neutrinos regardless of flavor. The electron fraction implied by these values is $Y_e \cong 0.13$, and the oscillation potentials are $\omega \cong 0.3$~km$^{-1}$, $\mu | \mathbf{D} (0) | \cong 3 \times 10^5$~km$^{-1}$, and $\lambda \cong 5 \times 10^7$~km$^{-1}$. The last of these is implemented in the calculation using a matter-suppressed mixing angle $\theta = 10^{-6}$ \cite{hannestad2006}. The mass hierarchy is not important for the results of this paper.

Crucially, this region is envisioned as being in the spatially extended atmosphere in which neutrinos decouple. NC scattering on neutrons is the dominant process affecting the heavy-lepton flavors. Using the fiducial values of the previous paragraph, the rate is estimated to be
\begin{equation}
\frac{1}{\lambda_{\nu n}} \sim \frac{n_n \sigma_0}{4} \left( \frac{1 + 3 g_A^2}{4} \right) \left( \frac{E_\nu}{m_e} \right)^2  \sim \frac{1}{1.93~\textrm{km}},
\end{equation}
where $n_n$ is the neutron density, $m_e$ is the electron rest mass, $g_A \cong -1.28$ is the axial-vector coupling constant, and $\sigma_0 = 4 G_F^2 m_e^2 / \pi$ \cite{burrows2006}. Corrections from inelasticity, recoil, and weak magnetism are ignored. NC scattering on protons is subdominant, and four-neutrino processes, $\nu\bar{\nu}$ annihilation to $e^+ e^-$, nucleon--nucleon bremsstrahlung, and the flavor-blind contribution from electron scattering are all calculated to be insignificant. While $\nu n$ scattering is included in the calculation producing Fig.~\ref{iso}, as expected it has no effect (because, again, this model is isotropic and single-energy). The important point is that, in this representative calculation, $\nu_x$ and $\bar{\nu}_x$ continue to scatter but are no longer chemically coupled.

In contrast, CC capture on nucleons remains relevant:
\begin{align}
\frac{1}{\lambda^{\textrm{abs}}_{\nu_e n}} &\sim n_n \sigma_0 \left( \frac{1+3g_A^2}{4} \right) \left( \frac{E_\nu + Q}{m_e} \right)^2  \left( 1 + 1.1 \frac{E_\nu}{m_n} \right) \notag \\
&\sim \frac{1}{0.417~\textrm{km}} \notag \\
\frac{1}{\lambda^{\textrm{abs}}_{\bar{\nu}_e p}} &\sim n_p \sigma_0 \left( \frac{1+3g_A^2}{4} \right) \left( \frac{E_\nu - Q}{m_e} \right)^2 \left( 1 - 7.1 \frac{E_\nu}{m_n} \right) \notag \\
&\sim \frac{1}{4.36~\textrm{km}},
\end{align}
where $Q = m_n - m_p$ and corrections from recoil and weak magnetism have been retained \cite{burrows2006}. These rates are the critical ones in the collisionally unstable evolution. Electron scattering is secondary to CC capture, but it makes the leading contributions to $\Gamma_{e,x}^{CC}$ and $\bar{\Gamma}_{e,x}^{CC}$. Using the cross sections \cite{bowers1982}
\begin{equation}
\sigma_{\nu_\alpha e} = \frac{3}{8} \sigma_0 c_{\nu_\alpha} \frac{\left( T + \frac{1}{4} \mu_e \right) E_\nu}{m_e^2}
\end{equation}
with $c_{\nu_e} \cong 2.21$, $c_{\bar{\nu}_e} \cong 0.93$, and $c_{\nu_x} = c_{\bar{\nu}_x} \cong 0.36$, the rates of (flavor-resolving) electron scattering are estimated to be
\begin{align}
\frac{1}{\lambda_{\nu_e e}} &\sim n_{e^-} \left( \sigma_{\nu_e e} - \sigma_{\nu_x e} \right) \sim \frac{1}{11.4~\textrm{km}} \notag \\
\frac{1}{\lambda_{\bar{\nu}_e e}} &\sim n_{e^-} \left( \sigma_{\bar{\nu}_e e} - \sigma_{\bar{\nu}_x e} \right) \sim \frac{1}{37.2~\textrm{km}}.
\end{align}

On the basis of Eq.~\eqref{steqtn} and the interaction rates, we expect Fig.~\ref{iso} to show flavor transformation on a timescale of $\mathcal{O}(10)$~$\mu$s. Indeed, collisional instability sets in just after $\sim 20$~$\mu$s, causing the $\nu_e$--$\nu_x$ coherence density to rise and the electron-flavor species to convert to the heavy-flavor ones (and vice versa). If $\Gamma$ and $\bar{\Gamma}$ are artificially set to a common value (\textit{e.g.}, the average of their actual values), oscillations are stable and strongly matter-suppressed, and the amount of flavor conversion is utterly negligible. Thus all of the flavor evolution seen in Fig.~\ref{iso} is due to the collisional instability associated with $\Gamma \neq \bar{\Gamma}$.

Following the initial conversion of flavor, $n_{\nu_e}$ and $n_{\bar{\nu}_e}$ return to their initial values due to ongoing emission from $e^{\pm}$ capture on nucleons. As this takes place, $n_{\nu_x}$ and $n_{\bar{\nu}_x}$ are essentially frozen in place because of their inefficient chemical coupling. The species-dependence of the collision rates is therefore doubly important. It not only destabilizes the flavor field but also differentially restores (classical) equilibrium. In effect, the conversion of $\nu_e$, $\bar{\nu}_e$ into $\nu_x$, $\bar{\nu}_x$ hides these particles from absorption, allowing for an overall enhancement of the neutrino luminosity.

As for the prevalence, we are now equipped to make several comments. The obvious statement is that collisional instabilities are only noteworthy in regions where neutrinos are not yet fully free-streaming. At the other extreme, they are unlikely to occur in neutrino-trapping regions, where degeneracy decrees a stabilizing hierarchy $n_{\nu_e} > n_{\nu_x} > n_{\bar{\nu}_e}$. The favored region has neutrinos \textit{partially} coupled to the medium, as in the fiducial case above.

\begin{figure}
\centering
\includegraphics[width=.43\textwidth]{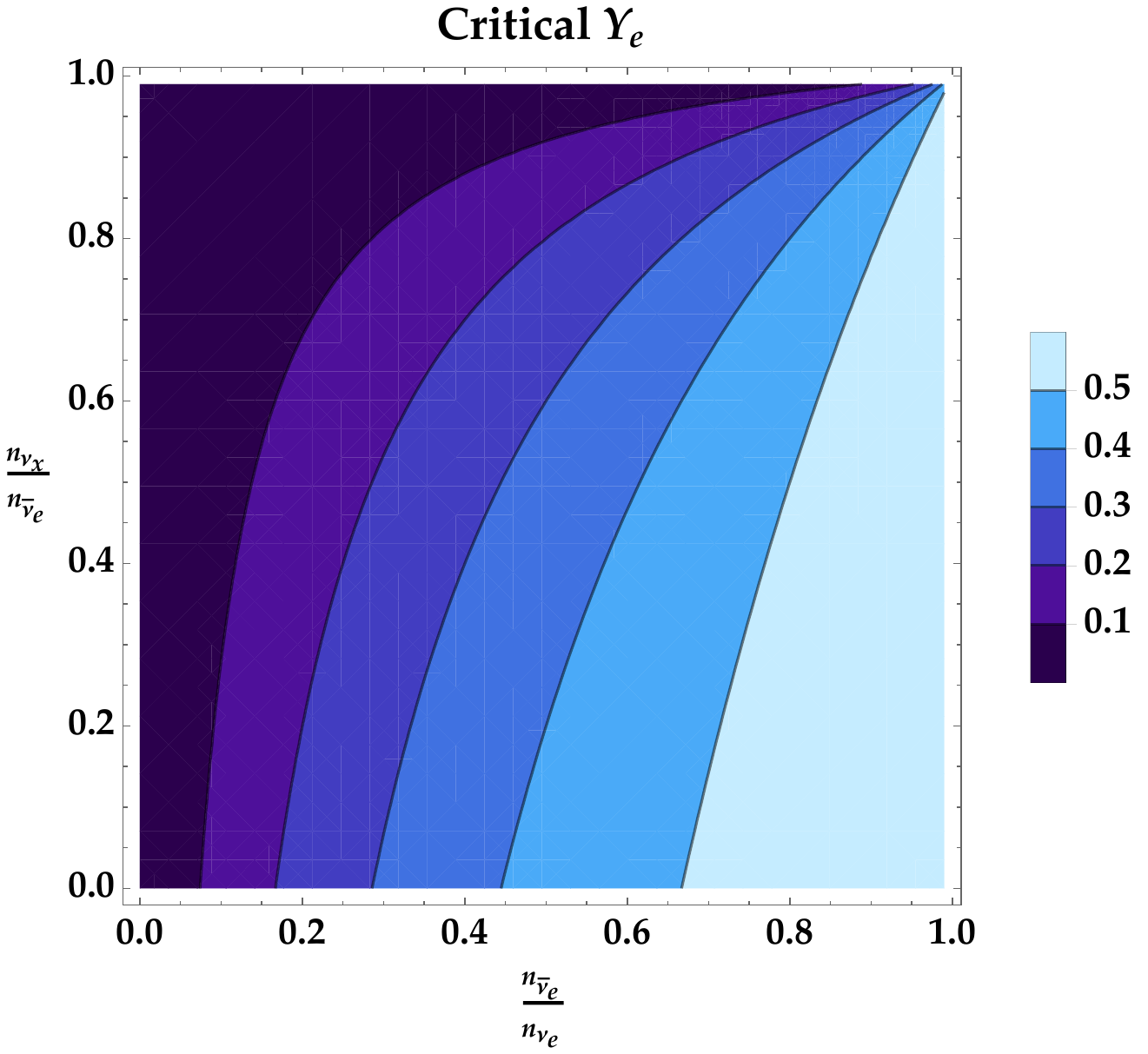}
\caption{The critical electron fraction $Y_e^\textrm{crit}$ below which the system is predicted to be collisionally unstable, shown as a function of $n_{\nu_x} / n_{\bar{\nu}_e}$ and $n_{\bar{\nu}_e} / n_{\nu_e}$ and assuming $n_{\nu_x} = n_{\bar{\nu}_x}$. Since $Y_e \lesssim 0.2$ is typical in the neutrino decoupling region, the majority of this parameter space is unstable.}
\label{critical_ye}
\end{figure}

To be more quantitative, where $\Gamma$ and $\bar{\Gamma}$ are well approximated by the rates of capture on nucleons, the critical electron fraction $Y_e^\textrm{crit}$ (such that $Y_e \lesssim Y_e^\textrm{crit}$ is unstable) can be found from the relation
\begin{equation}
Y_e^\textrm{crit} \cong \left( 1 + \frac{1}{\xi \mathcal{R}} \right)^{-1}. \label{yecrit}
\end{equation}
This follows from Eq.~\eqref{criterion} and $\lambda^\textrm{abs}_{\bar{\nu}_e p} / \lambda^\textrm{abs}_{\nu_e n} \cong \xi n_n / n_p$, where $\xi$ depends on the neutrino energy spectra. Fig.~\ref{critical_ye} shows $Y_e^\textrm{crit}$ with $\xi = 1.6$, consistent with the monochromatic approximation ($E_\nu = 20$~MeV) used throughout this paper. A more comprehensive version of this analysis could be done by specifying spectral information, but the conclusion is anticipated to be the same: collisional instability is likely to occur in some regions.

\textit{Extending the analysis.}---A curious feature of Eq.~\eqref{steqtn} is that it exhibits no dependence \textit{at all} on oscillation parameters. A system with $\omega = \mu = 0$ should therefore support the same solutions, assuming the initial state is seeded with flavor coherence. As a matter of fact, such a system does enter into the decay mode, but never into the growing one. From the vantage point of Eq.~\eqref{steqtn}, the significance of the oscillation terms is that they cause the polarization vectors to wander through different configurations in flavor space until chancing upon the growing solution. Fast instabilities, by way of contrast, really can arise with $\omega = 0$ as long as coherence is seeded. The $\mu$ terms serve double duty in those cases, prompting the exploration of flavor space and driving the instabilities themselves.

Linear stability analysis provides a complementary perspective. For this we return to the density matrices. Linearizing in off-diagonal elements and adopting a matter-suppressed mixing angle $\theta_m \cong 0$,
\begin{align}
i \partial_t \rho_{ex} &= \left( - \omega - \sqrt{2} G_F ( n_{\bar{\nu}_e} - n_{\bar{\nu}_x} )  - i \Gamma  \right) \rho_{ex} \notag \\
& \hspace{0.25 in} + \sqrt{2}G_F ( n_{\nu_e} - n_{\nu_x} ) \bar{\rho}_{ex} \notag \\
i \partial_t \bar{\rho}_{ex} &= \left( + \omega + \sqrt{2} G_F ( n_{\nu_e} - n_{\nu_x} )  - i \bar{\Gamma}  \right) \bar{\rho}_{ex} \notag \\
& \hspace{0.25 in} - \sqrt{2}G_F ( n_{\bar{\nu}_e} - n_{\bar{\nu}_x} ) \rho_{ex}. \label{rhoex}
\end{align}
Seeking collective modes, we now take $\rho_{ex} = Q e^{-i \Omega t}$ and $\bar{\rho}_{ex} = \bar{Q} e^{-i \Omega t}$. The dispersion relation results from plugging these expressions into Eqs.~\eqref{rhoex} and dispensing with $Q$ and $\bar{Q}$. It can be solved analytically:
\begin{equation}
\textrm{Im}~\Omega \cong \pm \frac{\Gamma - \bar{\Gamma}}{2} \frac{ \mu S }{\sqrt{(\mu D)^2 + 4 \omega \mu S}} - \frac{\Gamma + \bar{\Gamma}}{2}, \label{linstab}
\end{equation}
where $S = | \mathbf{S} (0) | = n_{\nu_e} - n_{\nu_x} + n_{\bar{\nu}_e} - n_{\bar{\nu}_x}$ and $D = | \mathbf{D} (0) | = n_{\nu_e} - n_{\nu_x} - n_{\bar{\nu}_e} + n_{\bar{\nu}_x}$. ($\mathbf{S}$ and $\mathbf{D}$ are assumed to point along $\mathbf{z}$ initially, but the formulas are easily adapted.) If $\mu D \gg 2\sqrt{\omega \mu S}$, which is usually expected of the setting we have in mind, then the instability criterion coincides with Eq.~\eqref{criterion}. If $\mu D < 2\sqrt{\omega \mu S}$ and $\omega < 0$ (indicating the inverted hierarchy), then Eq.~\eqref{linstab} is invalidated by intervention of the bipolar instability.

\begin{figure}
\centering
\includegraphics[width=.43\textwidth]{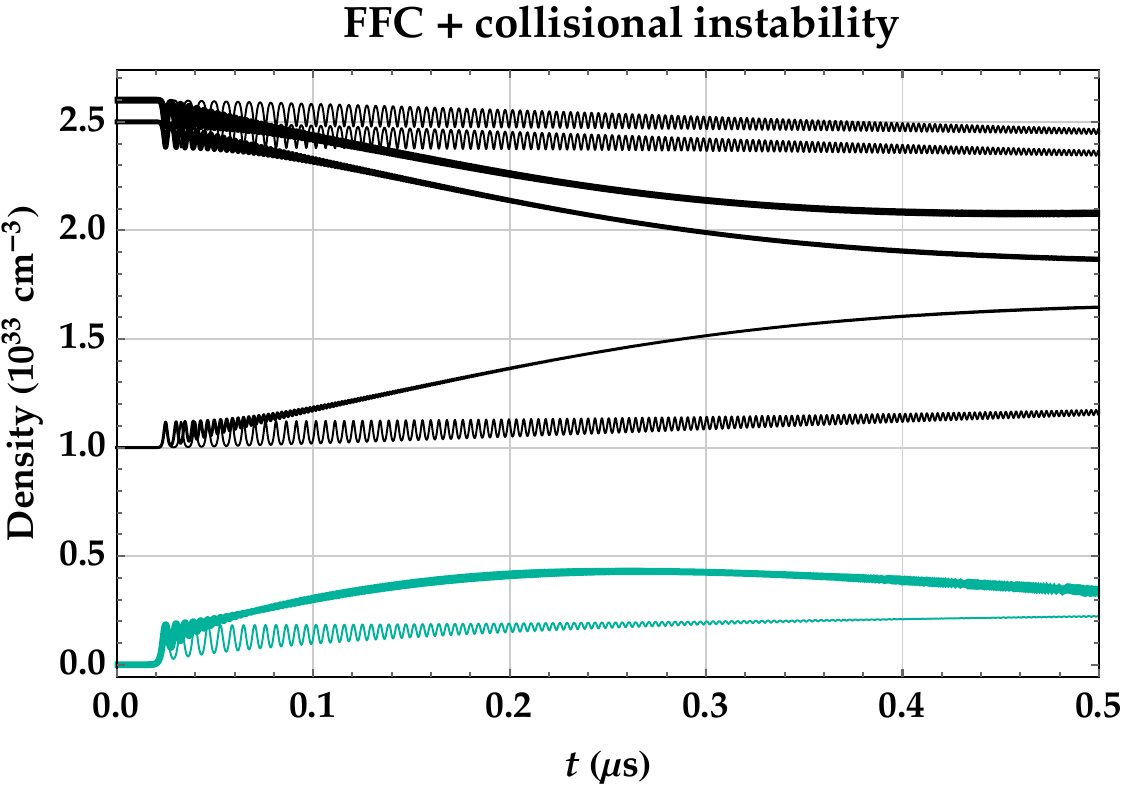}
\caption{Collisionally and fast-unstable evolution in an anisotropic calculation: $n_{\nu_e}$ (thick black curve), $n_{\bar{\nu}_e}$ (medium), $n_{\nu_x}$ (thin), and neutrino coherence density $| \mathbf{P}_T | / 2$ (teal). The very thin curves, which show only minor secular change, are the results when $\Gamma$ and $\bar{\Gamma}$ are artificially set to the average of their actual values (hence $\Gamma = \bar{\Gamma}$). The rapid oscillatory motion is the swinging of the fast pendulum \cite{johns2020}. No conversion would be visible if the system were stable to fast flavor conversion (FFC).}
\label{fast}
\end{figure}

Up to this point the analysis has assumed monochromaticity, isotropy, and homogeneity. The first of these is justified by the high neutrino density. Though not presented here, numerical calculations with multiple energies confirm that collisional instability affects them collectively.

Calculations also confirm the presence of collisionally unstable evolution in anisotropic set-ups. An interesting case is one where collisional and fast instabilities are present together. Fig.~\ref{fast} shows the results of such a calculation. The parameters are the same as those used in making Fig.~\ref{iso} except that $n_{\nu_e}$ has been decreased to $2.6 \times 10^{33}$~cm$^{-3}$ and the angular distributions have been made anisotropic, so as to make the system unstable to fast oscillations. As with the other parameters, the angular distributions are chosen to be representative of real conditions in a supernova. They are specified by the flux factors (\textit{i.e.}, the ratios of energy flux to energy density) $f_{\nu_e} = 0.05$,  $f_{\bar{\nu}_e} = 0.10$, and $f_{\nu_x} = f_{\bar{\nu}_x} = 0.15$. Radiative pressures are prescribed using M1 closure \cite{johns2021}. These distributions are nearly isotropic and are plotted in Fig.~1 of Ref.~\cite{johns2022}.

The onset of fast flavor conversion prompts the growth of the collisional instability on a much shorter timescale than was seen in Fig.~\ref{iso}. Furthermore, significantly greater flavor transformation occurs when $\Gamma \neq \bar{\Gamma}$ than when $\Gamma = \bar{\Gamma}$, testifying to the fact that the results observed in Fig.~\ref{fast} are not simply caused by decoherence. In a more realistic setting, collisional relaxation will compete with various forms of collisionless relaxation \cite{johns2020b, bhattacharyya2021} to determine the outcome. Nonetheless, the figure demonstrates that collisional instability has the potential to be enhanced by fast oscillations rather than wiped out by them.

Lastly, collisional instability is expected to occur in homogeneous and inhomogeneous environments alike, much as the bipolar instability is known to. In fact, preliminary evidence points to the collisional instability identified here---a homogeneous, isotropic, temporally growing mode---as one member of a family. These remarks need further development, however, and will be presented in a future publication.

\textit{Added note.}---Since the first version of this paper appeared, subsequent work has expanded on its major points \cite{johns2022b, padillagay2022b, lin2022, xiong2022, xiong2022b}.

\textit{Acknowledgements.}---Support for this work was provided by NASA through the NASA Hubble Fellowship grant number HST-HF2-51461.001-A awarded by the Space Telescope Science Institute, which is operated by the Association of Universities for Research in Astronomy, Incorporated, under NASA contract NAS5-26555.

\bibliography{all_papers}

\begin{thebibliography}{30}%
\makeatletter
\providecommand \@ifxundefined [1]{%
 \@ifx{#1\undefined}
}%
\providecommand \@ifnum [1]{%
 \ifnum #1\expandafter \@firstoftwo
 \else \expandafter \@secondoftwo
 \fi
}%
\providecommand \@ifx [1]{%
 \ifx #1\expandafter \@firstoftwo
 \else \expandafter \@secondoftwo
 \fi
}%
\providecommand \natexlab [1]{#1}%
\providecommand \enquote  [1]{``#1''}%
\providecommand \bibnamefont  [1]{#1}%
\providecommand \bibfnamefont [1]{#1}%
\providecommand \citenamefont [1]{#1}%
\providecommand \href@noop [0]{\@secondoftwo}%
\providecommand \href [0]{\begingroup \@sanitize@url \@href}%
\providecommand \@href[1]{\@@startlink{#1}\@@href}%
\providecommand \@@href[1]{\endgroup#1\@@endlink}%
\providecommand \@sanitize@url [0]{\catcode `\\12\catcode `\$12\catcode
  `\&12\catcode `\#12\catcode `\^12\catcode `\_12\catcode `\%12\relax}%
\providecommand \@@startlink[1]{}%
\providecommand \@@endlink[0]{}%
\providecommand \url  [0]{\begingroup\@sanitize@url \@url }%
\providecommand \@url [1]{\endgroup\@href {#1}{\urlprefix }}%
\providecommand \urlprefix  [0]{URL }%
\providecommand \Eprint [0]{\href }%
\providecommand \doibase [0]{https://doi.org/}%
\providecommand \selectlanguage [0]{\@gobble}%
\providecommand \bibinfo  [0]{\@secondoftwo}%
\providecommand \bibfield  [0]{\@secondoftwo}%
\providecommand \translation [1]{[#1]}%
\providecommand \BibitemOpen [0]{}%
\providecommand \bibitemStop [0]{}%
\providecommand \bibitemNoStop [0]{.\EOS\space}%
\providecommand \EOS [0]{\spacefactor3000\relax}%
\providecommand \BibitemShut  [1]{\csname bibitem#1\endcsname}%
\let\auto@bib@innerbib\@empty
\bibitem [{\citenamefont {Mirizzi}\ \emph {et~al.}(2016)\citenamefont
  {Mirizzi}, \citenamefont {Tamborra}, \citenamefont {Janka}, \citenamefont
  {Saviano}, \citenamefont {Scholberg}, \citenamefont {Bollig}, \citenamefont
  {Hudepohl},\ and\ \citenamefont {Chakraborty}}]{mirizzi2016}%
  \BibitemOpen
  \bibfield  {author} {\bibinfo {author} {\bibfnamefont {A.}~\bibnamefont
  {Mirizzi}}, \bibinfo {author} {\bibfnamefont {I.}~\bibnamefont {Tamborra}},
  \bibinfo {author} {\bibfnamefont {H.-T.}\ \bibnamefont {Janka}}, \bibinfo
  {author} {\bibfnamefont {N.}~\bibnamefont {Saviano}}, \bibinfo {author}
  {\bibfnamefont {K.}~\bibnamefont {Scholberg}}, \bibinfo {author}
  {\bibfnamefont {R.}~\bibnamefont {Bollig}}, \bibinfo {author} {\bibfnamefont
  {L.}~\bibnamefont {Hudepohl}},\ and\ \bibinfo {author} {\bibfnamefont
  {S.}~\bibnamefont {Chakraborty}},\ }\bibfield  {title} {\bibinfo {title}
  {{Supernova Neutrinos: Production, Oscillations and Detection}},\ }\href
  {https://doi.org/10.1393/ncr/i2016-10120-8} {\bibfield  {journal} {\bibinfo
  {journal} {Riv. Nuovo Cim.}\ }\textbf {\bibinfo {volume} {39}},\ \bibinfo
  {pages} {1} (\bibinfo {year} {2016})}\BibitemShut {NoStop}%
\bibitem [{\citenamefont {Banerjee}\ \emph {et~al.}(2011)\citenamefont
  {Banerjee}, \citenamefont {Dighe},\ and\ \citenamefont
  {Raffelt}}]{banerjee2011}%
  \BibitemOpen
  \bibfield  {author} {\bibinfo {author} {\bibfnamefont {A.}~\bibnamefont
  {Banerjee}}, \bibinfo {author} {\bibfnamefont {A.}~\bibnamefont {Dighe}},\
  and\ \bibinfo {author} {\bibfnamefont {G.}~\bibnamefont {Raffelt}},\
  }\bibfield  {title} {\bibinfo {title} {Linearized flavor-stability analysis
  of dense neutrino streams},\ }\href
  {https://doi.org/10.1103/PhysRevD.84.053013} {\bibfield  {journal} {\bibinfo
  {journal} {Phys. Rev. D}\ }\textbf {\bibinfo {volume} {84}},\ \bibinfo
  {pages} {053013} (\bibinfo {year} {2011})}\BibitemShut {NoStop}%
\bibitem [{\citenamefont {Izaguirre}\ \emph {et~al.}(2017)\citenamefont
  {Izaguirre}, \citenamefont {Raffelt},\ and\ \citenamefont
  {Tamborra}}]{izaguirre2017}%
  \BibitemOpen
  \bibfield  {author} {\bibinfo {author} {\bibfnamefont {I.}~\bibnamefont
  {Izaguirre}}, \bibinfo {author} {\bibfnamefont {G.}~\bibnamefont {Raffelt}},\
  and\ \bibinfo {author} {\bibfnamefont {I.}~\bibnamefont {Tamborra}},\
  }\bibfield  {title} {\bibinfo {title} {Fast pairwise conversion of supernova
  neutrinos: A dispersion relation approach},\ }\href
  {https://doi.org/10.1103/PhysRevLett.118.021101} {\bibfield  {journal}
  {\bibinfo  {journal} {Phys. Rev. Lett.}\ }\textbf {\bibinfo {volume} {118}},\
  \bibinfo {pages} {021101} (\bibinfo {year} {2017})}\BibitemShut {NoStop}%
\bibitem [{\citenamefont {Capozzi}\ \emph {et~al.}(2017)\citenamefont
  {Capozzi}, \citenamefont {Dasgupta}, \citenamefont {Lisi}, \citenamefont
  {Marrone},\ and\ \citenamefont {Mirizzi}}]{capozzi2017}%
  \BibitemOpen
  \bibfield  {author} {\bibinfo {author} {\bibfnamefont {F.}~\bibnamefont
  {Capozzi}}, \bibinfo {author} {\bibfnamefont {B.}~\bibnamefont {Dasgupta}},
  \bibinfo {author} {\bibfnamefont {E.}~\bibnamefont {Lisi}}, \bibinfo {author}
  {\bibfnamefont {A.}~\bibnamefont {Marrone}},\ and\ \bibinfo {author}
  {\bibfnamefont {A.}~\bibnamefont {Mirizzi}},\ }\bibfield  {title} {\bibinfo
  {title} {Fast flavor conversions of supernova neutrinos: Classifying
  instabilities via dispersion relations},\ }\href
  {https://doi.org/10.1103/PhysRevD.96.043016} {\bibfield  {journal} {\bibinfo
  {journal} {Phys. Rev. D}\ }\textbf {\bibinfo {volume} {96}},\ \bibinfo
  {pages} {043016} (\bibinfo {year} {2017})}\BibitemShut {NoStop}%
\bibitem [{\citenamefont {Sawyer}(2016)}]{sawyer2016}%
  \BibitemOpen
  \bibfield  {author} {\bibinfo {author} {\bibfnamefont {R.~F.}\ \bibnamefont
  {Sawyer}},\ }\bibfield  {title} {\bibinfo {title} {Neutrino cloud
  instabilities just above the neutrino sphere of a supernova},\ }\href
  {https://doi.org/10.1103/PhysRevLett.116.081101} {\bibfield  {journal}
  {\bibinfo  {journal} {Phys. Rev. Lett.}\ }\textbf {\bibinfo {volume} {116}},\
  \bibinfo {pages} {081101} (\bibinfo {year} {2016})}\BibitemShut {NoStop}%
\bibitem [{\citenamefont {Chakraborty}\ \emph {et~al.}(2016)\citenamefont
  {Chakraborty}, \citenamefont {Hansen}, \citenamefont {Izaguirre},\ and\
  \citenamefont {Raffelt}}]{chakraborty2016}%
  \BibitemOpen
  \bibfield  {author} {\bibinfo {author} {\bibfnamefont {S.}~\bibnamefont
  {Chakraborty}}, \bibinfo {author} {\bibfnamefont {R.}~\bibnamefont {Hansen}},
  \bibinfo {author} {\bibfnamefont {I.}~\bibnamefont {Izaguirre}},\ and\
  \bibinfo {author} {\bibfnamefont {G.}~\bibnamefont {Raffelt}},\ }\bibfield
  {title} {\bibinfo {title} {Collective neutrino flavor conversion: Recent
  developments},\ }\href
  {https://doi.org/http://dx.doi.org/10.1016/j.nuclphysb.2016.02.012}
  {\bibfield  {journal} {\bibinfo  {journal} {Nucl. Phys.}\ }\textbf {\bibinfo
  {volume} {B908}},\ \bibinfo {pages} {366 } (\bibinfo {year}
  {2016})}\BibitemShut {NoStop}%
\bibitem [{\citenamefont {Tamborra}\ and\ \citenamefont
  {Shalgar}(2020)}]{tamborra2020}%
  \BibitemOpen
  \bibfield  {author} {\bibinfo {author} {\bibfnamefont {I.}~\bibnamefont
  {Tamborra}}\ and\ \bibinfo {author} {\bibfnamefont {S.}~\bibnamefont
  {Shalgar}},\ }\bibfield  {title} {\bibinfo {title} {New developments in
  flavor evolution of a dense neutrino gas},\ }\href@noop {} {\bibfield
  {journal} {\bibinfo  {journal} {arXiv:2011.01948}\ } (\bibinfo {year}
  {2020})}\BibitemShut {NoStop}%
\bibitem [{\citenamefont {Duan}\ \emph {et~al.}(2010)\citenamefont {Duan},
  \citenamefont {Fuller},\ and\ \citenamefont {Qian}}]{duan2010}%
  \BibitemOpen
  \bibfield  {author} {\bibinfo {author} {\bibfnamefont {H.}~\bibnamefont
  {Duan}}, \bibinfo {author} {\bibfnamefont {G.~M.}\ \bibnamefont {Fuller}},\
  and\ \bibinfo {author} {\bibfnamefont {Y.-Z.}\ \bibnamefont {Qian}},\
  }\bibfield  {title} {\bibinfo {title} {Collective neutrino oscillations},\
  }\href@noop {} {\bibfield  {journal} {\bibinfo  {journal} {Ann. Rev. Nucl.
  Part. Sci.}\ }\textbf {\bibinfo {volume} {60}},\ \bibinfo {pages} {569}
  (\bibinfo {year} {2010})}\BibitemShut {NoStop}%
\bibitem [{\citenamefont {McKellar}\ and\ \citenamefont
  {Thomson}(1994)}]{mckellar1994}%
  \BibitemOpen
  \bibfield  {author} {\bibinfo {author} {\bibfnamefont {B.~H.~J.}\
  \bibnamefont {McKellar}}\ and\ \bibinfo {author} {\bibfnamefont {M.~J.}\
  \bibnamefont {Thomson}},\ }\bibfield  {title} {\bibinfo {title} {Oscillating
  neutrinos in the early universe},\ }\href@noop {} {\bibfield  {journal}
  {\bibinfo  {journal} {Phys. Rev. D}\ }\textbf {\bibinfo {volume} {49}},\
  \bibinfo {pages} {2710} (\bibinfo {year} {1994})}\BibitemShut {NoStop}%
\bibitem [{\citenamefont {Dolgov}(2001)}]{dolgov2001}%
  \BibitemOpen
  \bibfield  {author} {\bibinfo {author} {\bibfnamefont {A.~D.}\ \bibnamefont
  {Dolgov}},\ }\bibfield  {title} {\bibinfo {title} {Neutrino oscillations in
  the early universe. resonant case},\ }\href
  {https://doi.org/https://doi.org/10.1016/S0550-3213(01)00323-6} {\bibfield
  {journal} {\bibinfo  {journal} {Nucl. Phys.}\ }\textbf {\bibinfo {volume}
  {B610}},\ \bibinfo {pages} {411 } (\bibinfo {year} {2001})}\BibitemShut
  {NoStop}%
\bibitem [{\citenamefont {Hannestad}\ \emph {et~al.}(2015)\citenamefont
  {Hannestad}, \citenamefont {Hansen}, \citenamefont {Tram},\ and\
  \citenamefont {Wong}}]{hannestad2015}%
  \BibitemOpen
  \bibfield  {author} {\bibinfo {author} {\bibfnamefont {S.}~\bibnamefont
  {Hannestad}}, \bibinfo {author} {\bibfnamefont {R.~S.}\ \bibnamefont
  {Hansen}}, \bibinfo {author} {\bibfnamefont {T.}~\bibnamefont {Tram}},\ and\
  \bibinfo {author} {\bibfnamefont {Y.~Y.}\ \bibnamefont {Wong}},\ }\bibfield
  {title} {\bibinfo {title} {Active-sterile neutrino oscillations in the early
  universe with full collision terms},\ }\href
  {https://doi.org/10.1088/1475-7516/2015/08/019} {\bibfield  {journal}
  {\bibinfo  {journal} {J. Cosmol. Astropart. Phys.}\ }\textbf {\bibinfo
  {volume} {2015}}\bibinfo  {number} { (08)},\ \bibinfo {pages}
  {019}}\BibitemShut {NoStop}%
\bibitem [{\citenamefont {Johns}(2019)}]{johns2019b}%
  \BibitemOpen
\bibfield  {number} {  }\bibfield  {author} {\bibinfo {author} {\bibfnamefont
  {L.}~\bibnamefont {Johns}},\ }\bibfield  {title} {\bibinfo {title}
  {Derivation of the sterile neutrino boltzmann equation from quantum
  kinetics},\ }\href {https://doi.org/10.1103/PhysRevD.100.083536} {\bibfield
  {journal} {\bibinfo  {journal} {Phys. Rev. D}\ }\textbf {\bibinfo {volume}
  {100}},\ \bibinfo {pages} {083536} (\bibinfo {year} {2019})}\BibitemShut
  {NoStop}%
\bibitem [{\citenamefont {Richers}\ \emph {et~al.}(2019)\citenamefont
  {Richers}, \citenamefont {McLaughlin}, \citenamefont {Kneller},\ and\
  \citenamefont {Vlasenko}}]{richers2019}%
  \BibitemOpen
  \bibfield  {author} {\bibinfo {author} {\bibfnamefont {S.~A.}\ \bibnamefont
  {Richers}}, \bibinfo {author} {\bibfnamefont {G.~C.}\ \bibnamefont
  {McLaughlin}}, \bibinfo {author} {\bibfnamefont {J.~P.}\ \bibnamefont
  {Kneller}},\ and\ \bibinfo {author} {\bibfnamefont {A.}~\bibnamefont
  {Vlasenko}},\ }\bibfield  {title} {\bibinfo {title} {Neutrino quantum
  kinetics in compact objects},\ }\href
  {https://doi.org/10.1103/PhysRevD.99.123014} {\bibfield  {journal} {\bibinfo
  {journal} {Phys. Rev. D}\ }\textbf {\bibinfo {volume} {99}},\ \bibinfo
  {pages} {123014} (\bibinfo {year} {2019})}\BibitemShut {NoStop}%
\bibitem [{\citenamefont {Capozzi}\ \emph {et~al.}(2019)\citenamefont
  {Capozzi}, \citenamefont {Dasgupta}, \citenamefont {Mirizzi}, \citenamefont
  {Sen},\ and\ \citenamefont {Sigl}}]{capozzi2019}%
  \BibitemOpen
  \bibfield  {author} {\bibinfo {author} {\bibfnamefont {F.}~\bibnamefont
  {Capozzi}}, \bibinfo {author} {\bibfnamefont {B.}~\bibnamefont {Dasgupta}},
  \bibinfo {author} {\bibfnamefont {A.}~\bibnamefont {Mirizzi}}, \bibinfo
  {author} {\bibfnamefont {M.}~\bibnamefont {Sen}},\ and\ \bibinfo {author}
  {\bibfnamefont {G.}~\bibnamefont {Sigl}},\ }\bibfield  {title} {\bibinfo
  {title} {Collisional triggering of fast flavor conversions of supernova
  neutrinos},\ }\href {https://doi.org/10.1103/PhysRevLett.122.091101}
  {\bibfield  {journal} {\bibinfo  {journal} {Phys. Rev. Lett.}\ }\textbf
  {\bibinfo {volume} {122}},\ \bibinfo {pages} {091101} (\bibinfo {year}
  {2019})}\BibitemShut {NoStop}%
\bibitem [{\citenamefont {Shalgar}\ and\ \citenamefont
  {Tamborra}(2021)}]{shalgar2021}%
  \BibitemOpen
  \bibfield  {author} {\bibinfo {author} {\bibfnamefont {S.}~\bibnamefont
  {Shalgar}}\ and\ \bibinfo {author} {\bibfnamefont {I.}~\bibnamefont
  {Tamborra}},\ }\bibfield  {title} {\bibinfo {title} {Change of direction in
  pairwise neutrino conversion physics: The effect of collisions},\ }\href
  {https://doi.org/10.1103/PhysRevD.103.063002} {\bibfield  {journal} {\bibinfo
   {journal} {Phys. Rev. D}\ }\textbf {\bibinfo {volume} {103}},\ \bibinfo
  {pages} {063002} (\bibinfo {year} {2021})}\BibitemShut {NoStop}%
\bibitem [{\citenamefont {Martin}\ \emph {et~al.}(2021)\citenamefont {Martin},
  \citenamefont {Carlson}, \citenamefont {Cirigliano},\ and\ \citenamefont
  {Duan}}]{martin2021}%
  \BibitemOpen
  \bibfield  {author} {\bibinfo {author} {\bibfnamefont {J.~D.}\ \bibnamefont
  {Martin}}, \bibinfo {author} {\bibfnamefont {J.}~\bibnamefont {Carlson}},
  \bibinfo {author} {\bibfnamefont {V.}~\bibnamefont {Cirigliano}},\ and\
  \bibinfo {author} {\bibfnamefont {H.}~\bibnamefont {Duan}},\ }\bibfield
  {title} {\bibinfo {title} {Fast flavor oscillations in dense neutrino media
  with collisions},\ }\href {https://doi.org/10.1103/PhysRevD.103.063001}
  {\bibfield  {journal} {\bibinfo  {journal} {Phys. Rev. D}\ }\textbf {\bibinfo
  {volume} {103}},\ \bibinfo {pages} {063001} (\bibinfo {year}
  {2021})}\BibitemShut {NoStop}%
\bibitem [{\citenamefont {{Hannestad}}\ \emph {et~al.}(2006)\citenamefont
  {{Hannestad}}, \citenamefont {{Raffelt}}, \citenamefont {{Sigl}},\ and\
  \citenamefont {{Wong}}}]{hannestad2006}%
  \BibitemOpen
  \bibfield  {author} {\bibinfo {author} {\bibfnamefont {S.}~\bibnamefont
  {{Hannestad}}}, \bibinfo {author} {\bibfnamefont {G.~G.}\ \bibnamefont
  {{Raffelt}}}, \bibinfo {author} {\bibfnamefont {G.}~\bibnamefont {{Sigl}}},\
  and\ \bibinfo {author} {\bibfnamefont {Y.~Y.~Y.}\ \bibnamefont {{Wong}}},\
  }\bibfield  {title} {\bibinfo {title} {{Self-induced conversion in dense
  neutrino gases: Pendulum in flavor space}},\ }\href
  {https://doi.org/10.1103/PhysRevD.74.105010} {\bibfield  {journal} {\bibinfo
  {journal} {Phys. Rev. D}\ }\textbf {\bibinfo {volume} {74}},\ \bibinfo
  {pages} {105010} (\bibinfo {year} {2006})}\BibitemShut {NoStop}%
\bibitem [{\citenamefont {Johns}\ and\ \citenamefont
  {Fuller}(2018)}]{johns2018}%
  \BibitemOpen
  \bibfield  {author} {\bibinfo {author} {\bibfnamefont {L.}~\bibnamefont
  {Johns}}\ and\ \bibinfo {author} {\bibfnamefont {G.~M.}\ \bibnamefont
  {Fuller}},\ }\bibfield  {title} {\bibinfo {title} {Strange mechanics of the
  neutrino flavor pendulum},\ }\href
  {https://doi.org/10.1103/PhysRevD.97.023020} {\bibfield  {journal} {\bibinfo
  {journal} {Phys. Rev. D}\ }\textbf {\bibinfo {volume} {97}},\ \bibinfo
  {pages} {023020} (\bibinfo {year} {2018})}\BibitemShut {NoStop}%
\bibitem [{\citenamefont {Burrows}\ \emph {et~al.}(2006)\citenamefont
  {Burrows}, \citenamefont {Reddy},\ and\ \citenamefont
  {Thompson}}]{burrows2006}%
  \BibitemOpen
  \bibfield  {author} {\bibinfo {author} {\bibfnamefont {A.}~\bibnamefont
  {Burrows}}, \bibinfo {author} {\bibfnamefont {S.}~\bibnamefont {Reddy}},\
  and\ \bibinfo {author} {\bibfnamefont {T.~A.}\ \bibnamefont {Thompson}},\
  }\bibfield  {title} {\bibinfo {title} {Neutrino opacities in nuclear
  matter},\ }\href
  {https://doi.org/https://doi.org/10.1016/j.nuclphysa.2004.06.012} {\bibfield
  {journal} {\bibinfo  {journal} {Nucl. Phys. A}\ }\textbf {\bibinfo {volume}
  {777}},\ \bibinfo {pages} {356 } (\bibinfo {year} {2006})},\ \bibinfo {note}
  {special Issue on Nuclear Astrophysics}\BibitemShut {NoStop}%
\bibitem [{\citenamefont {Bowers}\ and\ \citenamefont
  {Wilson}(1982)}]{bowers1982}%
  \BibitemOpen
  \bibfield  {author} {\bibinfo {author} {\bibfnamefont {R.~L.}\ \bibnamefont
  {Bowers}}\ and\ \bibinfo {author} {\bibfnamefont {J.~R.}\ \bibnamefont
  {Wilson}},\ }\bibfield  {title} {\bibinfo {title} {A numerical model for
  stellar core collapse calculations},\ }\href@noop {} {\bibfield  {journal}
  {\bibinfo  {journal} {Astrophys. J. Suppl. Ser.}\ }\textbf {\bibinfo {volume}
  {50}},\ \bibinfo {pages} {115} (\bibinfo {year} {1982})}\BibitemShut
  {NoStop}%
\bibitem [{\citenamefont {Johns}\ \emph
  {et~al.}(2020{\natexlab{a}})\citenamefont {Johns}, \citenamefont {Nagakura},
  \citenamefont {Fuller},\ and\ \citenamefont {Burrows}}]{johns2020}%
  \BibitemOpen
  \bibfield  {author} {\bibinfo {author} {\bibfnamefont {L.}~\bibnamefont
  {Johns}}, \bibinfo {author} {\bibfnamefont {H.}~\bibnamefont {Nagakura}},
  \bibinfo {author} {\bibfnamefont {G.~M.}\ \bibnamefont {Fuller}},\ and\
  \bibinfo {author} {\bibfnamefont {A.}~\bibnamefont {Burrows}},\ }\bibfield
  {title} {\bibinfo {title} {Neutrino oscillations in supernovae: Angular
  moments and fast instabilities},\ }\href
  {https://doi.org/10.1103/PhysRevD.101.043009} {\bibfield  {journal} {\bibinfo
   {journal} {Phys. Rev. D}\ }\textbf {\bibinfo {volume} {101}},\ \bibinfo
  {pages} {043009} (\bibinfo {year} {2020}{\natexlab{a}})}\BibitemShut
  {NoStop}%
\bibitem [{\citenamefont {Johns}\ and\ \citenamefont
  {Nagakura}(2021)}]{johns2021}%
  \BibitemOpen
  \bibfield  {author} {\bibinfo {author} {\bibfnamefont {L.}~\bibnamefont
  {Johns}}\ and\ \bibinfo {author} {\bibfnamefont {H.}~\bibnamefont
  {Nagakura}},\ }\bibfield  {title} {\bibinfo {title} {{Fast flavor
  instabilities and the search for neutrino angular crossings}},\ }\href@noop
  {} {\  (\bibinfo {year} {2021})},\ \Eprint {https://arxiv.org/abs/2104.04106}
  {arXiv:2104.04106 [hep-ph]} \BibitemShut {NoStop}%
\bibitem [{\citenamefont {Johns}\ and\ \citenamefont
  {Nagakura}(2022)}]{johns2022}%
  \BibitemOpen
  \bibfield  {author} {\bibinfo {author} {\bibfnamefont {L.}~\bibnamefont
  {Johns}}\ and\ \bibinfo {author} {\bibfnamefont {H.}~\bibnamefont
  {Nagakura}},\ }\href@noop {} {\bibinfo {title} {Self-consistency in models of
  neutrino scattering and fast flavor conversion}} (\bibinfo {year} {2022}),\
  \Eprint {https://arxiv.org/abs/2206.09225} {arXiv:2206.09225 [hep-ph]}
  \BibitemShut {NoStop}%
\bibitem [{\citenamefont {Johns}\ \emph
  {et~al.}(2020{\natexlab{b}})\citenamefont {Johns}, \citenamefont {Nagakura},
  \citenamefont {Fuller},\ and\ \citenamefont {Burrows}}]{johns2020b}%
  \BibitemOpen
  \bibfield  {author} {\bibinfo {author} {\bibfnamefont {L.}~\bibnamefont
  {Johns}}, \bibinfo {author} {\bibfnamefont {H.}~\bibnamefont {Nagakura}},
  \bibinfo {author} {\bibfnamefont {G.~M.}\ \bibnamefont {Fuller}},\ and\
  \bibinfo {author} {\bibfnamefont {A.}~\bibnamefont {Burrows}},\ }\bibfield
  {title} {\bibinfo {title} {Fast oscillations, collisionless relaxation, and
  spurious evolution of supernova neutrino flavor},\ }\href
  {https://doi.org/10.1103/PhysRevD.102.103017} {\bibfield  {journal} {\bibinfo
   {journal} {Phys. Rev. D}\ }\textbf {\bibinfo {volume} {102}},\ \bibinfo
  {pages} {103017} (\bibinfo {year} {2020}{\natexlab{b}})}\BibitemShut
  {NoStop}%
\bibitem [{\citenamefont {Bhattacharyya}\ and\ \citenamefont
  {Dasgupta}(2021)}]{bhattacharyya2021}%
  \BibitemOpen
  \bibfield  {author} {\bibinfo {author} {\bibfnamefont {S.}~\bibnamefont
  {Bhattacharyya}}\ and\ \bibinfo {author} {\bibfnamefont {B.}~\bibnamefont
  {Dasgupta}},\ }\bibfield  {title} {\bibinfo {title} {Fast flavor
  depolarization of supernova neutrinos},\ }\href
  {https://doi.org/10.1103/PhysRevLett.126.061302} {\bibfield  {journal}
  {\bibinfo  {journal} {Phys. Rev. Lett.}\ }\textbf {\bibinfo {volume} {126}},\
  \bibinfo {pages} {061302} (\bibinfo {year} {2021})}\BibitemShut {NoStop}%
\bibitem [{\citenamefont {Johns}\ and\ \citenamefont
  {Xiong}(2022)}]{johns2022b}%
  \BibitemOpen
  \bibfield  {author} {\bibinfo {author} {\bibfnamefont {L.}~\bibnamefont
  {Johns}}\ and\ \bibinfo {author} {\bibfnamefont {Z.}~\bibnamefont {Xiong}},\
  }\bibfield  {title} {\bibinfo {title} {Collisional instabilities of neutrinos
  and their interplay with fast flavor conversion in compact objects},\ }\href
  {https://doi.org/10.1103/PhysRevD.106.103029} {\bibfield  {journal} {\bibinfo
   {journal} {Phys. Rev. D}\ }\textbf {\bibinfo {volume} {106}},\ \bibinfo
  {pages} {103029} (\bibinfo {year} {2022})}\BibitemShut {NoStop}%
\bibitem [{\citenamefont {Padilla-Gay}\ \emph {et~al.}(2022)\citenamefont
  {Padilla-Gay}, \citenamefont {Tamborra},\ and\ \citenamefont
  {Raffelt}}]{padillagay2022b}%
  \BibitemOpen
  \bibfield  {author} {\bibinfo {author} {\bibfnamefont {I.}~\bibnamefont
  {Padilla-Gay}}, \bibinfo {author} {\bibfnamefont {I.}~\bibnamefont
  {Tamborra}},\ and\ \bibinfo {author} {\bibfnamefont {G.~G.}\ \bibnamefont
  {Raffelt}},\ }\bibfield  {title} {\bibinfo {title} {Neutrino fast flavor
  pendulum. ii. collisional damping},\ }\href
  {https://doi.org/10.1103/PhysRevD.106.103031} {\bibfield  {journal} {\bibinfo
   {journal} {Phys. Rev. D}\ }\textbf {\bibinfo {volume} {106}},\ \bibinfo
  {pages} {103031} (\bibinfo {year} {2022})}\BibitemShut {NoStop}%
\bibitem [{\citenamefont {Lin}\ and\ \citenamefont {Duan}(2022)}]{lin2022}%
  \BibitemOpen
  \bibfield  {author} {\bibinfo {author} {\bibfnamefont {Y.-C.}\ \bibnamefont
  {Lin}}\ and\ \bibinfo {author} {\bibfnamefont {H.}~\bibnamefont {Duan}},\
  }\bibfield  {title} {\bibinfo {title} {Collision-induced flavor instability
  in dense neutrino gases with energy-dependent scattering},\ }\href@noop {}
  {\bibfield  {journal} {\bibinfo  {journal} {arXiv preprint arXiv:2210.09218}\
  } (\bibinfo {year} {2022})}\BibitemShut {NoStop}%
\bibitem [{\citenamefont {Xiong}\ \emph
  {et~al.}(2022{\natexlab{a}})\citenamefont {Xiong}, \citenamefont {Wu},
  \citenamefont {Mart{\'\i}nez-Pinedo}, \citenamefont {Fischer}, \citenamefont
  {George}, \citenamefont {Lin},\ and\ \citenamefont {Johns}}]{xiong2022}%
  \BibitemOpen
  \bibfield  {author} {\bibinfo {author} {\bibfnamefont {Z.}~\bibnamefont
  {Xiong}}, \bibinfo {author} {\bibfnamefont {M.-R.}\ \bibnamefont {Wu}},
  \bibinfo {author} {\bibfnamefont {G.}~\bibnamefont {Mart{\'\i}nez-Pinedo}},
  \bibinfo {author} {\bibfnamefont {T.}~\bibnamefont {Fischer}}, \bibinfo
  {author} {\bibfnamefont {M.}~\bibnamefont {George}}, \bibinfo {author}
  {\bibfnamefont {C.-Y.}\ \bibnamefont {Lin}},\ and\ \bibinfo {author}
  {\bibfnamefont {L.}~\bibnamefont {Johns}},\ }\bibfield  {title} {\bibinfo
  {title} {Evolution of collisional neutrino flavor instabilities in
  spherically symmetric supernova models},\ }\href@noop {} {\bibfield
  {journal} {\bibinfo  {journal} {arXiv preprint arXiv:2210.08254}\ } (\bibinfo
  {year} {2022}{\natexlab{a}})}\BibitemShut {NoStop}%
\bibitem [{\citenamefont {Xiong}\ \emph
  {et~al.}(2022{\natexlab{b}})\citenamefont {Xiong}, \citenamefont {Johns},
  \citenamefont {Wu},\ and\ \citenamefont {Duan}}]{xiong2022b}%
  \BibitemOpen
  \bibfield  {author} {\bibinfo {author} {\bibfnamefont {Z.}~\bibnamefont
  {Xiong}}, \bibinfo {author} {\bibfnamefont {L.}~\bibnamefont {Johns}},
  \bibinfo {author} {\bibfnamefont {M.-R.}\ \bibnamefont {Wu}},\ and\ \bibinfo
  {author} {\bibfnamefont {H.}~\bibnamefont {Duan}},\ }\bibfield  {title}
  {\bibinfo {title} {Collisional flavor instability in dense neutrino gases},\
  }\href@noop {} {\bibfield  {journal} {\bibinfo  {journal} {arXiv preprint
  arXiv:2212.03750}\ } (\bibinfo {year} {2022}{\natexlab{b}})}\BibitemShut
  {NoStop}%
\end{thebibliography}%

\end{document}